\documentclass[usenatbib]{mn2e}
\usepackage{fixltx2e,longtable,ifpdf,afterpage}
\ifpdf
  \usepackage{aeguill}
  \usepackage[pdftex]{graphicx,color}
\else
  \usepackage[dvips]{graphicx,color}
\fi

\newcommand\2{{\sc{ii}}}
\newcommand\1{{\sc{i}}}

\newcommand{\kms}{{\ensuremath{\mbox{km}\;\mbox{s}^{-1}}}}

\newcommand{\pA}{\ensuremath{\phantom{1}}}

\title[Kinematics of massive stars in the SMC]
{Kinematics of massive stars in the Small Magellanic Cloud}

\author[C. J.~Evans \& I. D.~Howarth]
{Christopher J.~Evans$^1$\thanks{email: cje@roe.ac.uk, idh@star.ucl.ac.uk}
\& Ian D.~Howarth$^2$
\\
$^1$UK Astronomy Technology Centre, Royal Observatory Edinburgh, 
Blackford Hill, Edinburgh, EH9 3HJ,~UK \\
$^2$Department of Physics \& Astronomy, University
College London, Gower Street, London WC1E~6BT,~UK \\
}
\date{Accepted by MNRAS}

\begin{document}
\maketitle

\begin{abstract}
We present radial velocities for 2045 stars in the Small Magellanic
Cloud (SMC), obtained from the 2dF survey by \citet{eh04}.  The great
majority of these stars are of OBA type, tracing the dynamics of the
young stellar population.  Dividing the sample into {\it ad hoc} `bar'
and `wing' samples (north and south, respectively, of the line:
$\delta = -77^{\circ}50^{\prime} + [4\alpha]^{\prime}$, where $\alpha$
is in minutes of time) we find that the velocities in the SMC bar show
a gradient of $26.3 \pm 1.6$~\kms~deg$^{-1}$ at a position angle of
$126 \pm 4^\circ$.  The derived gradient in the bar is robust to the
adopted line of demarcation between the two samples.  The largest
redshifts are found in the SMC wing, in which the velocity
distribution appears distinct from that in the bar, most likely a
consequence of the interaction between the Magellanic Clouds that is
predicted to have occurred 0.2~Gyr ago.  The mean velocity for all
stars in the sample is $+172.0 \pm 0.2$~\kms (redshifted by
$\sim$20~\kms\/ when compared to published results for older
populations), with a velocity dispersion of 30~\kms.
\end{abstract}


\section{Introduction}
\label{intro}

By studying the kinematic and chemical characteristics of the Milky
Way and nearby galaxies, past interactions and the history of mass
assembly can be explored.  The Magellanic Clouds are of particular
interest in this context as their dynamical evolution and
star-formation histories appear to be closely entwined
\citep[e.g.,][]{yn03}.  For instance, the Magellanic Stream has
generally been thought to arise from tidal disruption by the Milky Way
\citep{gn96,put98,ckg06}, although new proper motion results \citep{kma06}
suggest that the Clouds may be on their first passage about the Milky
Way, calling into question the origins of the Stream \citep{b07}.

From observations of the neutral-hydrogen gas, \citet{mf84} advanced
the concept that the SMC is comprised of two components, with a
`Mini-Magellanic Cloud' being effectively torn off by a past
interaction with the LMC, leaving the remnant of the SMC.  These
components had been identified previously by \citet{mn81}, who also
noted additional features at larger and smaller velocities.
Observations of the H$\;$\1 gas at better spatial resolution \citep{ss97}
have revealed a much more complex structure, comprised of hundreds of
expanding shells, presumably driven by stellar-wind outflows and
supernovae, which led the authors to describe the SMC as `distinctly
frothy and filamentary'.  Subsequent observations of three large supershells
were reported by \citet{stan99}, with the most dominant of these, 304A, potentially
accounting for the twin components seen previously in the main body of
the SMC.

To date, the most comprehensive survey of stellar kinematics in
the SMC is the study of 2046 red giants by \citet{hz06}, who concluded
that this old population is supported by its velocity dispersion
(rather than rotation).  Previous, more limited, studies have focussed
on luminous supergiants \citep{am77,am79,maa87}, Cepheids
\citep{dsm88}, and on the older stellar populations, e.g., carbon
stars \citep{dh97,kdi00} and red-clump stars \citep{hch93}.

The proximity and significantly sub-solar metallicity of the Clouds
have also made them popular targets for spectroscopic studies of
stellar evolution.  This was the primary motivation for our own
spectroscopic survey of the young, blue population of the Small
Magellanic Cloud \citep[][hereinafter Paper~I]{eh04}, which resulted
in spectral classifications for 4161 stars, dominated by early (OBA)
types.  Here we present radial velocities for 2045 objects from that
sample, covering a relatively wide spatial area (including the `wing'
region) and providing kinematic information for a component of the SMC
population which has previously gone largely unexplored.

\begin{figure*}
\begin{center}
\includegraphics[scale=0.65]{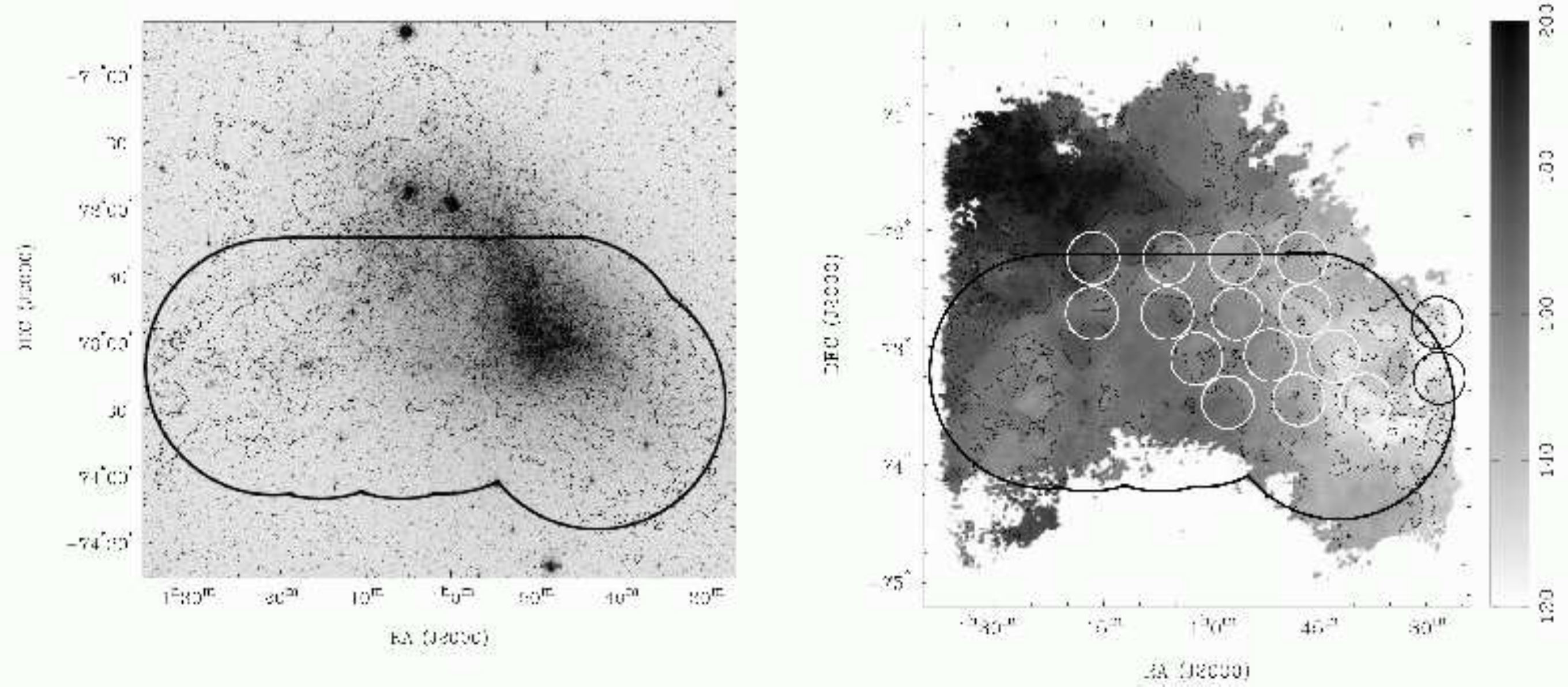}
\caption[]{The approximate boundaries of the 2dF SMC survey,
  superimposed on a $V$-band image by M.~Bessel (left; contours show
  H$\;$\1\ column densities), and on an H$\;$\1 velocity map (right;
  small circles show fields observed by \citealt{hz06}).  The dense
  optical region constitutes the SMC bar, with the Wing being
  the lower-density eastern region.  Adapted, with permission, from
  \citet{stan04}.}
\label{fig:stan}
\end{center}
\end{figure*}

\section{Stellar radial velocities}
\label{velocities}

Our spectroscopic survey used the 2-degree-field instrument at the
Anglo-Australian Telescope (2dF at the AAT).  2dF was a multi-fibre
instrument which allowed up to 400 intermediate-dispersion spectra to
be recorded simultaneously, using two spectrographs \citep{lewis02}.
The main observing programme consisted of 18 fields observed in two
runs: 1998 September and 1999 September (fields 1--12 and 13--18,
respectively). The survey area is shown in Fig.~\ref{fig:stan}.

Our spectra all cover at least the wavelength range
$\sim$3900--4800{\AA} (with about half including H$\beta$) at a
resolving power of $R \simeq 1600$ ($\Delta{v} \simeq 190$~\kms), with
signal-to-noise ratios in the range 20--150 (see Paper~I for details).
While the data were not optimized for very precise studies of
velocities for individual stars, the moderately large sample, and
extensive spatial coverage, encouraged us to investigate global trends
in the overall kinematics.

The 4161 observed targets cover a moderately wide range in spectral type
(with classifications from mid O-type to, in a few cases, late G-type),
and hence also in temperature and in spectral morphology.  Although we
compiled radial-velocity measurements using semi-automated procedures,
we found that, as a consequence of this range and of variations in
data quality, `manual' measurements gave better results.  Compelled
to use strong lines because of signal-to-noise constraints, we measured
heliocentric velocities for H$\beta$, H$\gamma$, H$\delta$, H8, and H9
(H$\epsilon$ was excluded due to blending with Ca~\2 $H$).
It is these results that we report
here.

Many of the spectra show emission in the lower members of the Balmer
series, precluding reliable measurement of the line centres.  In the
majority of cases this emission is probably nebular in origin, but in
some it will be intrinsic to the star.  Affected spectra were excluded
from the outset, leaving 3369 stars with three or more
`good' lines.

\subsection{External sources of error}
\label{sec:err}
Our data are free from one common source of radial-velocity error,
namely mis-centring of the target in the slit, because of `scrambling'
of the signal by the fibre runs.
However, the two 2dF spectrographs were mounted on the top-end ring of
the AAT, bringing other external factors, such as flexure,
into the radial-velocity
uncertainties. Such factors are well known to 2dF users (though not well
documented), with a signature of obvious problems being offsets in the velocity
zero-points between the two spectrographs (e.g., the unpublished
technical report by Stanford \& Cannon
2002\footnote{http://www.aao.gov.au/AAO/2df/technotes/fibvel.ps.gz}).

\begin{figure*}
\begin{center}
\includegraphics[scale=0.9]{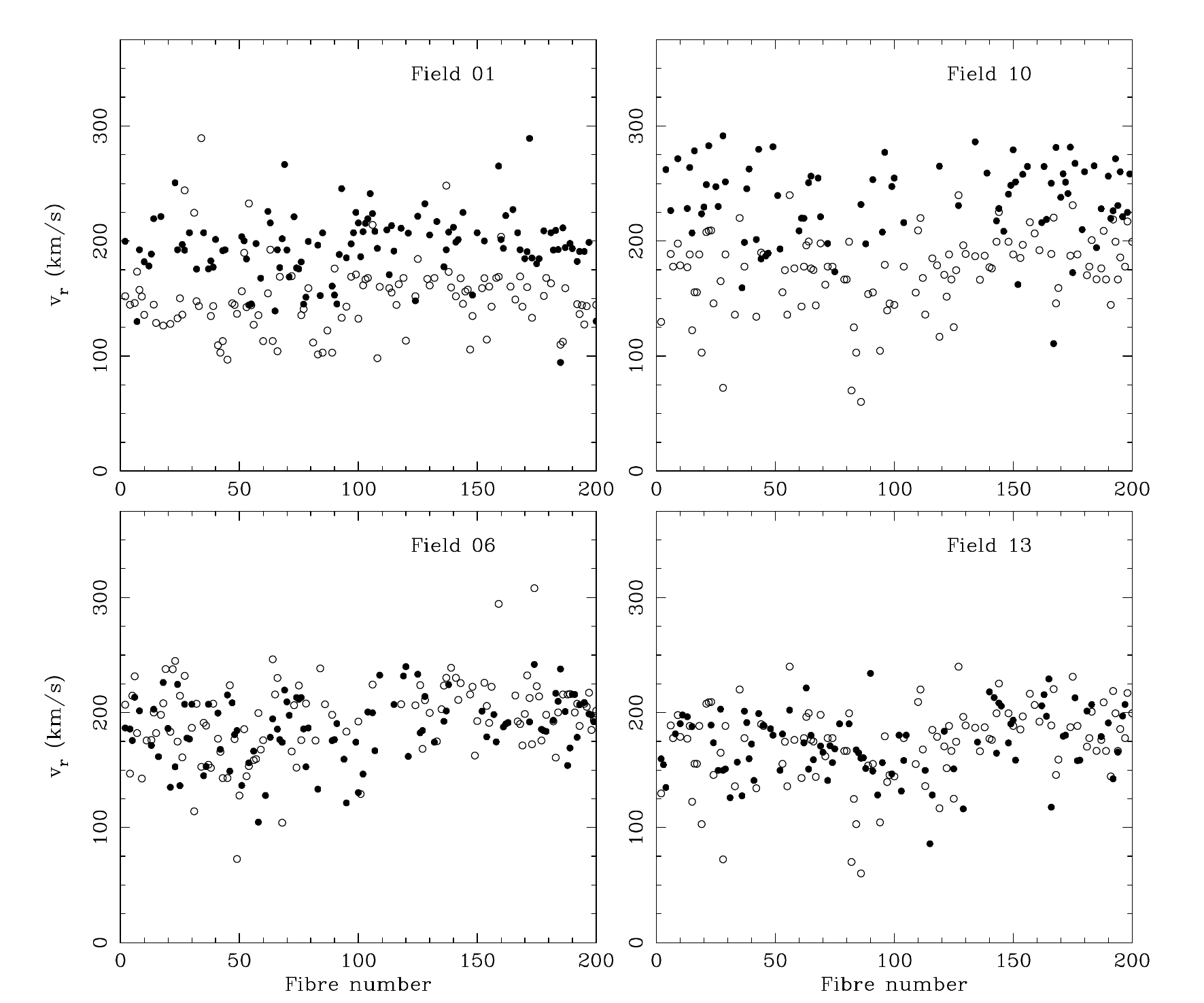}
\caption[]{Measured radial velocities from survey fields
1, 6, 10 \& 13.  Solid circles are velocities from spectrograph 1, open
circles those from spectrograph 2.  Fields 6 and 13 show no important
systematic offset; Field 1 and 10 show larger differences (median
velocities differ by 47 and 81~\kms, respectively), and were excluded
from the analysis.}
\label{fibrechecks}
\end{center}
\end{figure*}

Velocities from the two spectrographs appear well matched in our 
data throughout 1999 (as judged from comparison of the median velocities from the two
sets of different targets observed simultaneously at a given telescope
pointing).  However, the 1998 observations show poorer agreement, with
offsets of up to $\sim$100~\kms\ between spectrographs
(Fig.~\ref{fibrechecks}).  We rejected all fields where median
velocities from the two spectrographs differ by more than 30~\kms,
leaving results from 11 fields (3, 5, 6, 8, 9, 13--18) with a mean
offset (Spec1$-$Spec2) of $+3.1 \pm 14.1$ (s.d.)~\kms.  (A cut at
20~\kms\ further excludes fields 8 and 9, but has no material affect
on any results.)

We also used measurements of 66 stars from ancillary spectra
obtained in 2002; although these were taken in only one spectrograph,
the observations (described by \citealt{hhh03}) were designed for
radial-velocity measurements, and repeat observations show a scatter
consistent with internal errors alone.  This gives a final sample of
2045 stars, for which results are tabulated in Appendix~A.

\subsection{Internal uncertainties}
\label{sec:inchk}
Data given in the Appendix include the standard error on the mean of the
measurements for each star (i.e. $\sigma$/$\sqrt{n}$, where $n$ is the
number of measured lines).  The mean standard error from 2045 stars is
11.1~\kms, which provides an estimate of the purely internal
uncertainty of our velocities.  Only $\sim$3\%\ of stars have standard
errors greater than 25~\kms, and these can generally be attributed
to low signal-to-noise spectra, or to unrecognized nebular contamination.

Velocities are available from two separate 2dF observations for only
24 targets from the 2045-star sample; results are shown in
Fig.~\ref{2dfv2df} (which includes, for completeness, results for an
additional 14 stars with observations from rejected fields).  The
pairs of observations are distributed among 14 field/spectrograph
combinations, so that it is impractical to try to establish separately
any systematic offsets from these data.  Instead, we simply compute
the mean absolute difference between observations of each target:
$12.4 \pm 2.1$ (s.e.)~\kms.  It can be shown that the implied
1-$\sigma$ uncertainty on a single observation is a factor 0.886
smaller; i.e., $11.0 \pm 1.9$~\kms. This `internal error' estimate
incorporates any contributions from plate-to-plate zero-point shifts
and astrophysical variability in radial velocity, in addition to
stochastic uncertainties in the measurements; the excellent agreement
with the value given in the preceding paragraph therefore implies that
stochastic uncertainties dominate.

\subsection{External checks}
\label{sec:exchk}

\begin{figure}
\begin{center}
\includegraphics{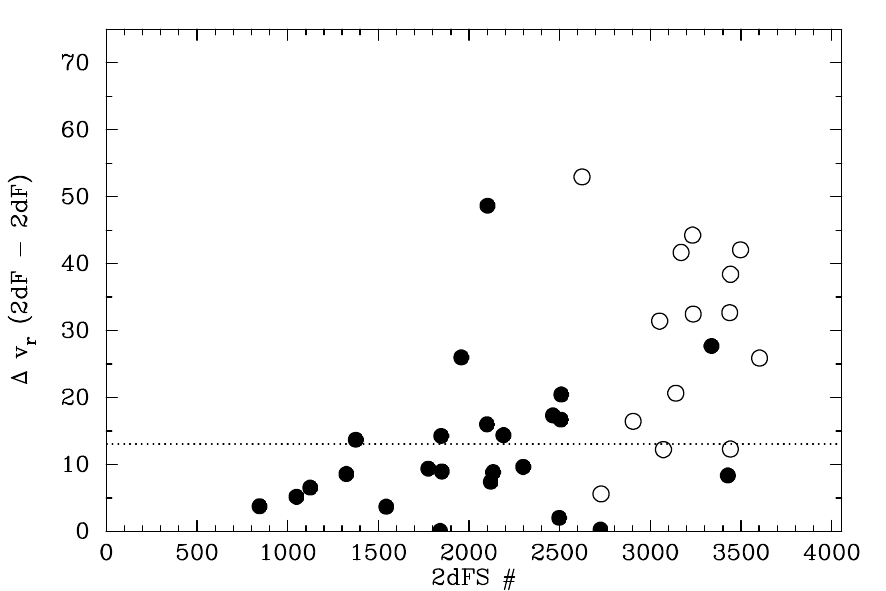}
\caption[]{Differences in radial velocities from repeat observations
  with 2dF.  Open symbols indicate stars from rejected fields
  (Section~\ref{sec:err}); the outlying filled symbol (2dFS~2102) may
  indicate a spectroscopic binary.  The horizontal dotted line is drawn at
  13.0~\kms, the internal 1-$\sigma$ error estimated from these data
  (Section~\ref{sec:inchk}).}
\label{2dfv2df}

\includegraphics{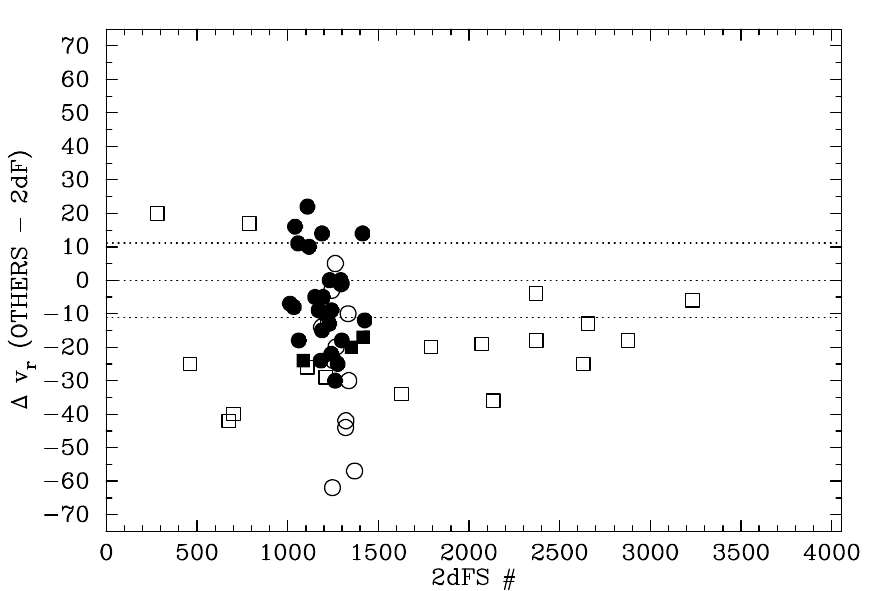}
\caption[]{Comparison of radial velocities from observations with
high-resolution VLT-FLAMES spectra \citep[filled circles,][]{flames2};
low-resolution VLT-FLAMES spectra (open circles, Evans et al., in
preparation); RGO spectra \citep[open squares,][]{eh03}; other
published high-resolution spectra (filled squares; see
Section~\ref{sec:exchk} for sources).  }
\label{2dfvflames}

\end{center}
\end{figure}

Of independent results at our disposal, the high-resolution
($R\sim$20,000, $\Delta{v} \simeq 15$~\kms) VLT-FLAMES survey of
massive stars in two SMC fields by \citet{flames2} offers both the
largest overlap with the 2dF data, and the highest-quality data.  A
comparison between 2dF and FLAMES velocities for 28 apparently single
stars is shown in Fig.~\ref{2dfvflames}.  There is excellent overall
agreement, with a mean offset of $-7$~\kms (FLAMES$-$2dF).  The
standard errors of the FLAMES velocities are small (typically
2-3~\kms); if the errors add quadratically, and if none of the stars
has intrinsic radial-velocity variations, the dispersion in
differential velocities (14~\kms) provides an upper limit to the
statistical uncertainty of the 2dF results of 13.5~\kms.

We have provisional velocities for a further 13 stars from a follow-up
FLAMES study of the SMC cluster NGC\,346, at lower resolution
($R\sim$7,000, $\Delta{v} \simeq 42$~\kms; Evans et al., in
preparation).  Results for these stars are are in reasonable agreement
with the 2dF values (Fig.~\ref{2dfvflames}), excepting four O9.5--B1
stars which have FLAMES velocities that are rather smaller than 2dF
values (by $\sim$40--70~\kms); these stars lead to a relatively large
dispersion in the velocity differences.  It is not implausible that these could all be
single-lined binaries (even though there is only a 1 in 8 probability
of all four stars having velocity offsets in the same direction).

Differential velocities are also shown in Fig.~\ref{2dfvflames} for 18
A-type supergiants/bright giants observed with the AAT's RGO spectrograph
at $R\sim$2,700 \citep{eh03}; in these relatively cooler, less massive
stars, effects such as nebular contamination and binarity would not be
expected to be as important as in the OB samples.  The 2dF
measurements show a $\sim$20~\kms\ offset (in the same sense as found
for the FLAMES comparison), which is only $\sim$10\%\
of the combined resolution elements of the two datasets.  If we
suppose that the dispersion in velocity differences between the 2dF
and RGO spectra arises from these datasets in inverse proportion to
their resolutions, the implied statistical uncertainty on 2dF
measurements is $\sim$15~\kms.

We also found three other apparently good-quality velocities in
the recent literature for seemingly single stars in common with the
2dF sample.  
These are
AzV\,170
\citep[][2dFS\#1086]{wal00},
AzV\,215 (\citealt{tl04},\footnote{\citeauthor{tl04} (\citeyear{tl04};
also \citealt{tl05}) label tabulated velocities as `$v_{\rm lsr}$',
but values are actually heliocentric; Dr.~C.~Trundle, personal
communication.} 2dFS\#1352) and AzV\,235 \citep[][2dFS\#1416, where
`2dFS' numbers are from the catalogue of Paper~I]{ecfh}.  Results are
included in Fig.~\ref{2dfvflames}.

The results of all these velocity comparisons are summarized in
Table~\ref{rv_diff}.  All the indications are that the 1-$\sigma$
statistical uncertainties in our 2dF velocities are
$\sim13 \pm 2$~\kms.  There is some evidence that the 2dF results may
be systematically too positive, by perhaps $\sim$10~\kms, but there is
no suggestion that this offset, if real, is a function of time or
position.  Given the general level of internal consistency demonstrated by
Figs.~\ref{2dfv2df} and~\ref{2dfvflames}, we are confident
that our results are, at the least, internally robust.

\begin{table}
\begin{center}
  \caption{Comparison of 2dF radial velocities from repeat exposures,
    and with those from external sources (discussed in
    Section~\ref{sec:exchk}). Tabulated velocity differences are in
    the sense `source' \textit{minus} 2dF.
  All targets used in the comparisons are OB
    stars, except for the RGO-spectrograph sample which is comprised
    of A-type stars.\label{rv_diff}}
\begin{tabular}{lrcccc}
\hline
Source & 
\multicolumn{1}{c}{$R$}& $N$ & $\overline{\Delta(v_{r})}$ & $\sigma$ & Median \\
            &&                &           & [\kms]      &   \\
\hline
2dF & 1600&24 &         & 13 &         \\
FLAMES (HR) & 20\,000& 28 & \pA$-$7 & 14 & \pA$-$9 \\
FLAMES (LR) & 7000&13 & $-$25 & 21 & $-$20 \\
RGO & 2700&18 & $-$18 & 17 & $-$20 \\
`Other' & 20\,000&\pA{3} & $-$19    & (4) & $-$18 \\
\hline
\end{tabular}
\end{center}
\end{table}

\section{Results}
\label{sec:vgrad}

\begin{figure*}
\begin{center}
\includegraphics[scale=0.6]{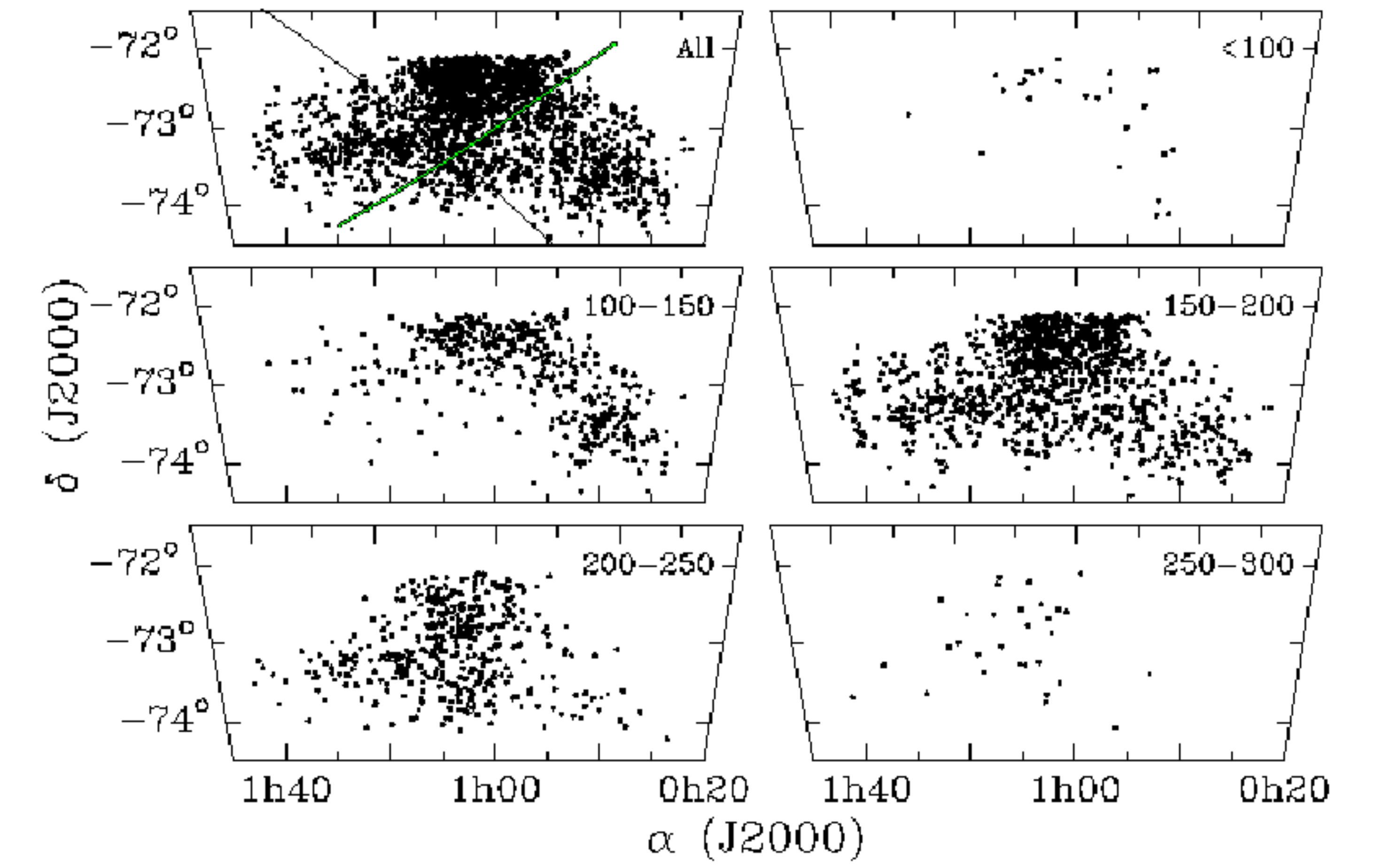}
\caption[]{Spatial distribution of the 2045 stars with radial velocities
from the 2dF survey.  The upper-left panel shows the whole sample,
with the black line dividing the sample into {\it ad hoc} subsets of
wing and bar stars (Section~\ref{sec:vgrad}).  
The axis of the maximum velocity gradient of the
bar stars, for a centre at 01h~00m $-73^\circ$~00$\arcmin$, is shown
in green (light grey in the printed version).  The other panels show
the positions of stars in 50~\kms~bins.}
\label{fig:vrplots}
\end{center}
\end{figure*}

\begin{figure}
\begin{center}
\includegraphics[scale=1.0]{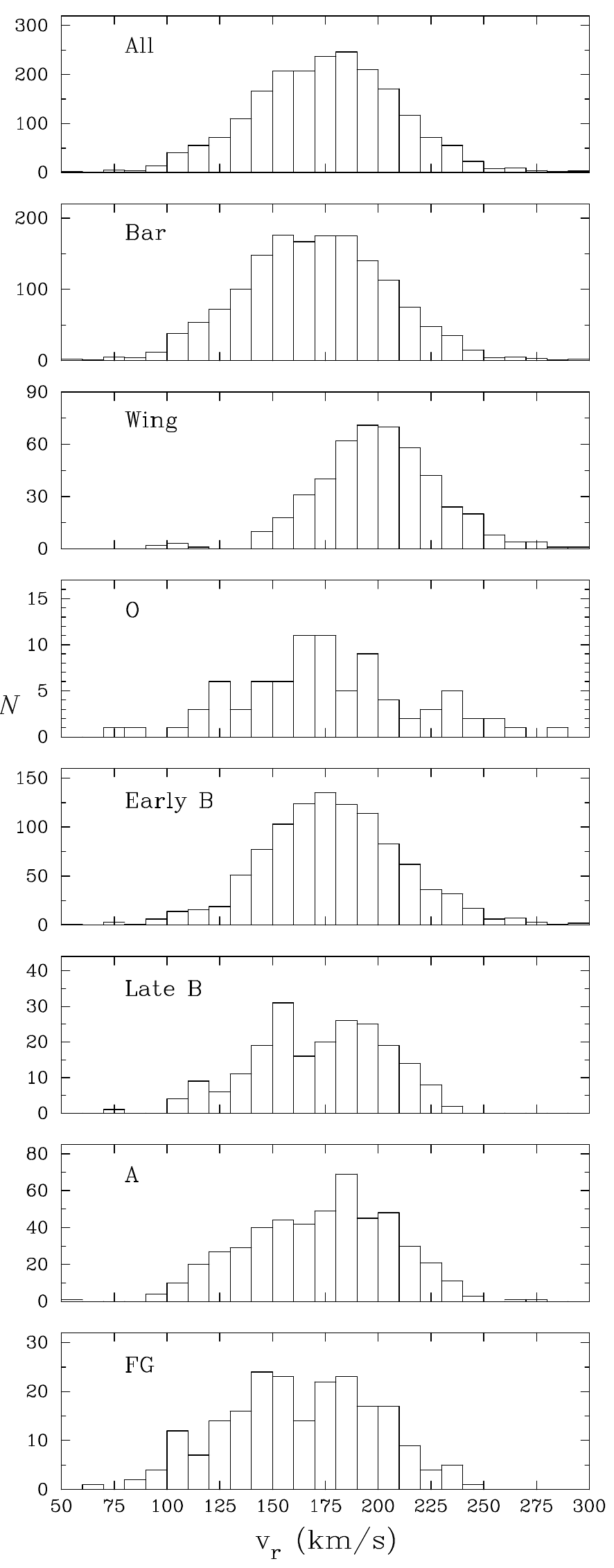}
\caption[]{Velocity histograms of the 2dF results for the bar and wing sub-samples, 
and as a function of spectral type.}
\label{fig:hist}
\end{center}
\end{figure}

Figure~\ref{fig:vrplots} shows the spatial distribution of our sample
by velocity bin.  By inspection, there is a trend for more positive
velocities to the south and east in our sample; in particular, the
wing region appears to have systematically more positive velocities
than the bar.  Arbitrarily dividing the sample stars into `bar' and
`wing' according to whether they lie north or south of a line:
\[ 
\delta = -77^{\circ}50^{\prime} + (4\alpha)^{\prime}
\] 
(Fig.~\ref{fig:vrplots}, where $\alpha$ is the right ascension in
minutes of time), histograms of the 2dF velocities for the two regions
are shown in Figure~\ref{fig:hist}.  Also shown are the results
grouped by spectral type; B-type spectra are grouped as either `Early
B' (earlier than B5) or `Late B' (B5, B8 or B9), with the 14 peculiar
AF-type stars included with the FG stars.

The most striking feature in Figure~\ref{fig:hist} is the difference
between the velocity distribution of the early B-type stars compared
with the later types.  The early B-type distribution is
near-symmetrical and single-peaked, centred on 175~\kms.  This is in
strong contrast to the FG stars which appear to have a
double-peaked distribution, with maxima at 150 and 180~\kms.  This
internal comparison reveals different velocity components in the
different spectral groups, i.e. with age\footnote{The appearance of
the distributions is, of course, a factor of the adopted bin-size.
For convenience in Figure~\ref{fig:hist} we chose 10~\kms.  Adopting a
bin-size of 15~\kms\/ still reveals a clear difference between the
early B- and FG-type results, although the two maxima are less
pronounced in the latter distribution.}; the relatively small number
of O-type stars are also peaked at $\sim$175~\kms.  The late B-type
results appear to have a similar distribution to the FG group, whereas
the picture is less clear for the A-type stars.

At first glance the 2dF results appear at odds with the velocities of
OBA stars from \citet{maa87}, in which a double-peaked structure is
seen, with the high-velocity peak much larger.  However, their
velocities \citep[which include the measurements by][]{am77,am79} are
predominantly from late B- and A-type stars which, when compared to
our results in Figure~\ref{fig:hist}, could account for the
double-peaked distribution.  Moreover, the remainder of the
\citeauthor{maa87} sample comprises 23 O- and early B-type stars, of
which 17 have positive Local Standard of Rest velocities (consistent
with our O- and early B-type distributions in Figure~\ref{fig:hist}),
giving the larger, high-velocity peak in their results.

One of the more interesting structures reported by
\citet{stan99,stan04} is the supergiant shell (SGS) 304A.  This spans
most of the northern region of our observations, with almost 800 of
the 2dF stars with radial velocities within the ellipse traced out by
the shell \citep[Table 2,][]{stan99}.  Over half of the stars
are O- and early B-type stars, and their velocity distribution is
indistinguishable from those in Figure~\ref{fig:hist}.  The velocity
diagram for SGS~304A shows two distinct components at approximately
130 and 170~\kms \citep{stan99}, presumably the origin of the offset
components reported by \citet{mn81} and \citet{mf84}.  
There is significant H$\alpha$ emission at the edge of SGS~304A
\citep[Figure~11,][]{stan99}, with N66
(NGC\,346, the largest H~\2 region in the SMC) at the northern limit.
Indeed, the higher-velocity component is dominant in the R.A.-velocity
slice at $-72^\circ 13\arcmin 40\arcsec$ \citep[Figure 6,][]{stan04}.
It appears that the majority of recent star formation in this region,
as traced by the 2dF early-type stars, is associated with the
higher-velocity component.

The FG stars observed with 2dF must be intrinsically luminous for
them if at the distance of the SMC.  Detailed luminosity
classification of these later-type spectra was beyond the scope
of Paper~I, however at this point it is of interest to consider their
physical luminosities, in order to estimate the range of ages and initial masses
probed by the 2dF results.
The median magnitude of the observed FG stars, the majority of which
are classified as F-type (187 from 229), is $B = 15.9$.  Adopting
$(B-V)_0 = 0.37$ and a bolometric correction of $0.11^{\rm m}$ for a
F5 supergiant \citep[e.g.][]{j66}, together with an SMC reddening of
$E(B-V) = 0.09$ \citep{pm95} and a distance modulus of 18.91
\citep{hhh05}, gives $M_{\rm bol} \sim -3.5$.  Evolutionary tracks
from \citet{char93} then suggest an initial mass of
$\sim$5M$_{\odot}$, (corresponding to a main-sequence spectral type of
$\sim$B2--3) and an age of $\sim$90--100 Myr.  These are only rough
estimates, but they demonstrate that the FG stars are beginning to
probe a population significantly older than that traced by the
early-type, main-sequence spectra.

\subsection{Mean velocities}
\label{sec:vdisp1}
The mean velocities and weighted dispersions, by spectral type, are
summarized in Table~\ref{tab:vels}.  The OBA-star velocities are
significantly more redshifted than results both for our relatively
small `late-type' (FG) sample, and for many other older-star samples
in the literature, which cluster around $\sim$150~\kms\ (see the
summary in Table~3 of \citealt{hz06}).  The latter difference arises
in part because our spatial coverage extends further to the east than
many previous studies, which, coupled with the velocity gradient we
have found, introduces a positive bias.  In particular, \citet{hz06}
report $\overline{v} = 145.6 \pm 0.6$~\kms\ for a red-giant population
with a mean position about a degree west of that for our sample
(Fig.~\ref{fig:stan}), where the mean velocity of the OBA stars is
$\sim$160~\kms.  While the residual difference is certainly within the
combined systematic uncertainties in the two datasets (and is of the
size and sign discussed in Section~\ref{sec:exchk}), it is of interest
to note that the mean of the $\gamma$ velocities determined for 50
eclipsing-binary OB stars by \citet{hhh03}\footnote{Heliocentric
corrections were applied in the wrong sense by Harries et al. (2003;
cf. Hilditch et al. 2005); we have corrected their $\gamma$ velocities
by +11~\kms\ to allow for this.} and \citet{hhh05} is $+196.2 \pm 24.8$
(s.d.)~\kms.  Their sample is, roughly, spatially coincident with
\citeauthor{hz06}'s, but is a full 50~\kms\ more redshifted on
average.

\begin{table}
\caption{Weighted mean velocities and 1$\sigma$ standard deviations 
of the 2dF sample, grouped by spectral classification.}
\label{tab:vels}
\begin{center}
\begin{tabular}{lrcc}
\hline
Type & $N\;$ & $\overline{v}$ & $\sigma$ \\
& & [\kms] & [\kms] \\
\hline
  O  &  83   &  $174.1   \pm   0.9 $&41.2\\
  B  &1251   &  $175.2   \pm   0.2 $&32.1\\
  A  & 482   &  $170.9   \pm   0.4 $&33.2\\
  FG & 229   &  $160.8   \pm   0.5 $&35.1\\
All  & 2045   &  $172.0   \pm   0.2 $&33.6\\
Bar & 1572 & $167.4  \pm  0.2$&33.3\\
Wing & 473 & $189.5  \pm  0.4$&28.6\\
\hline
\end{tabular}
\end{center}
\end{table}

The difference for the OB stars is in the opposite sense to that which
might be expected if the Balmer-line velocities were affected by
stellar-wind outflows (which should, if anything, give rise to a
blueshifted bias).  It remains to be seen how much of this is a
consequence of prosaic effects (e.g, unrecognized line blends, or
systematic offsets in 2dF spectroscopy), as compared to truly
different dynamics for the massive-star population.

\subsubsection{Velocities in the K1 region}
The K1 region lies to the east of the SMC bar  \citep{vf72}; it is included in
our wing sub-sample.  This region contains the N83 and N84 emission
complexes \citep{hen56}, and has been noted in the past as having a larger
velocity of recession than both the bar and wing populations
\citep[e.g.][]{am79,mml89}.  We also find larger velocities for the 2dF stars in this region:
$\overline{v} = 199.4 \pm 1.7$, $\sigma = 35.6$, compared to
$\overline{v} = 189.0 \pm 0.4$, $\sigma = 28.1$ for the remaining 446
stars in our wing sample, where we have adopted a radius of
7.5$\arcmin$ centred on N84A \citep[$\alpha = 1{\rm h} 14{\rm
m} 37{\rm s,}~\delta = -73^\circ 18\arcmin 27\arcsec$;][]{mml89,bd00}.

\subsection{Global gradients}
\label{sec:global}

\begin{figure*}
\begin{center}
\includegraphics[scale=0.6,angle=-90]{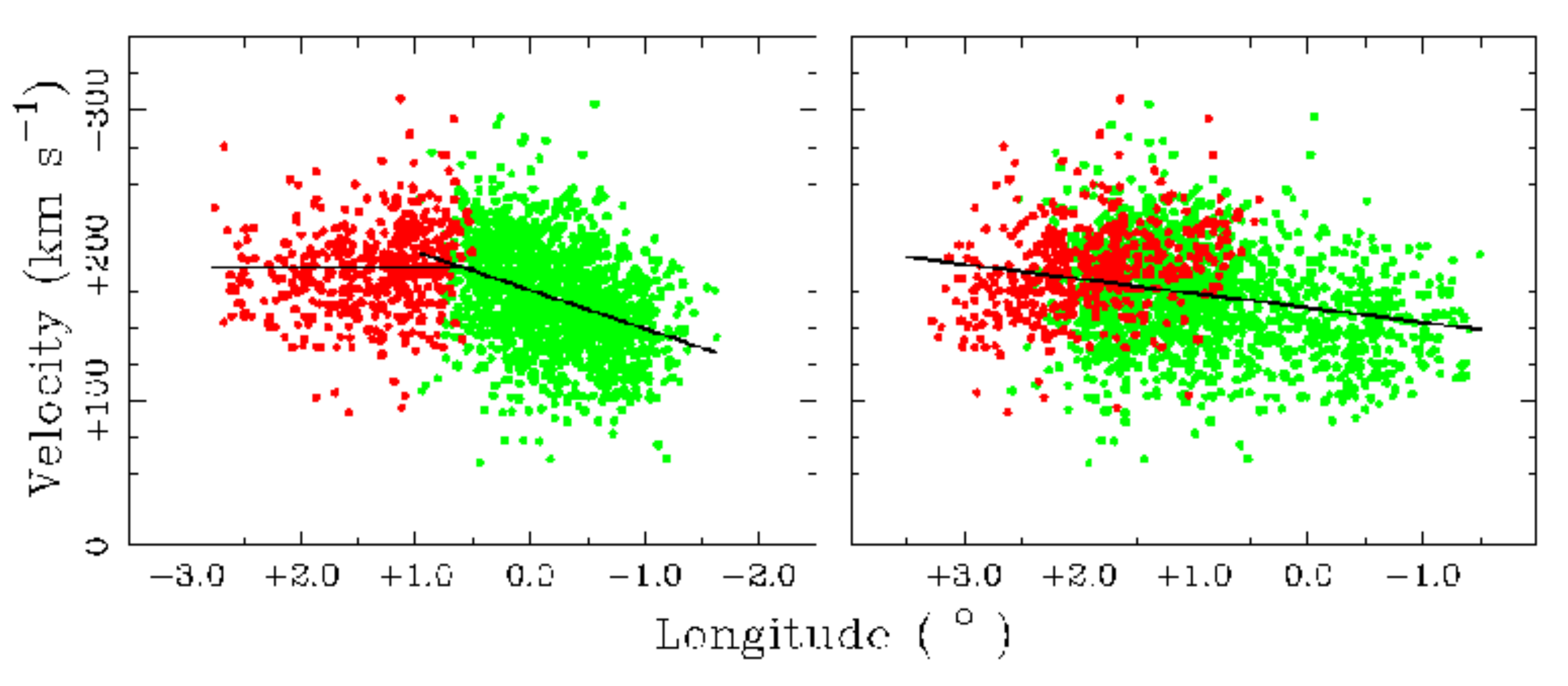}
\caption[]{{\em Left:} projection of radial velocities about 01h~00m,
  $-73^\circ$~0$\arcmin$ at PA~=~126$^\circ$; stars assigned to the
  SMC bar (Section~\ref{sec:vgrad}) are shown in green (right-hand
  grouping, light grey in the printed version), with those in the wing
  shown in red (darker grey).  The solid lines show the gradient in
  the bar ($26.5$~\kms~deg$^{-1}$), and the relatively flat gradient
  further east, in the wing, as discussed in Section~\ref{sec:vgrad}.
  {\em Right:} projection of radial velocities about 0h~45m,
  $-73^\circ$~0$\arcmin$ at PA~=~60$^\circ$, the direction of maximum
  gradient in H$\;$\1\ velocities (Section~\ref{sec:vgrad}); the line
  has a slope of 10~\kms~deg$^{-1}$.  }\label{fig:rotcurv}
\end{center}
\end{figure*}

To provide a simple quantification of the global velocity gradients in
our dataset (Figure~\ref{fig:vrplots}) we took two approaches.
First, as a reasonably nonparametric treatment, we set up a series of
spherical-co\"ordinate systems, each defined by a right ascension and
declination of zero latitude \& longitude (increasing north and east,
respectively), with the equatorial great circle rotated by position
angle $\theta$ (measured E from N) from the line of right ascension
passing through the central point.  For each such system we determined
the coefficient of correlation\footnote{We computed the Pearson
product-moment correlation coefficient $r$, Spearman's $\rho$
statistic, and Kendall's $\tau$; all give sensibly the same results.}
between stellar longitude and radial velocity; where the correlation
coefficient $r$ reaches a maximum, the maximum velocity gradient lies
along the equator.

Denoting the position angle of maximum velocity gradient as
$\theta_{\rm max}$, we find that the null hypothesis of no velocity
gradient can be formally rejected with extremely high confidence
($>99.999$\%) at $\theta_{\rm max}$, for any chosen central point, with
\[\theta_{\rm max} \simeq 126^\circ + 0.24(50-\alpha) \] 
(where $\alpha$ is again the right ascension of the adopted central point in
minutes of time, and the optical centre of the SMC is near $\alpha =
50$m) -- roughly perpendicular to the major axis of optical (and
H$\;$\1) emission.  The uncertainty in $\theta_{\rm max}$ is of order
5$^\circ$ (see below), while the small variations in $r(\theta_{\rm
  max})$ do not usefully constrain the location of the dynamical
centre.

As a more straighforward alternative, we also fit
velocities as a simple linear function of position,
\[
v = v_0 + A\Delta\alpha + B\Delta\delta,
\]
where $\Delta\alpha$, $\Delta\delta$ are co-ordinate offsets in degrees,
measured in cylindrical equidistant projection.   This is just a
plane, with $v_0$ the velocity of the plane at the origin of the
co-ordinate system.   Using the full dataset gives
a velocity gradient of $14.8 \pm 1.0$~\kms~deg$^{-1}$
at PA~=~$127.3 \pm 4.6^\circ$, with $v_0 = +173.47 \pm 0.75$~\kms\
at $1^{\rm h}.0, -73^\circ.0$.

Figure~\ref{fig:rotcurv} shows velocities in a co\"ordinate system
centred on 01h~00m, $-73^\circ$~00$\arcmin$ along PA~=~126$^\circ$. The
velocity gradient here averages 14.2~\kms~deg$^{-1}$ globally, but
evidently most of the velocity change is across the SMC bar.
Adopting the previous definition of `bar' and `wing' samples, the 473 stars in the
wing show roughly constant velocity (averaging +189.5~\kms), while the
bar stars show a gradient of $26.3 \pm 1.6$~\kms~deg$^{-1}$ 
at PA~=~$126.2 \pm 3.9^\circ$ about a mean velocity of 167.4~\kms.
Note that this result is robust to the details of the adopted bar/wing 
separation, as shown in Table~\ref{tab:grads}.

\begin{table*}
\caption{Velocity gradients derived from 2dF velocities in the SMC bar, as defined by stars north
of the lines given in the first column (where $\alpha$ is the right ascension in minutes
of time).  Small differences are found depending on the adopted
bar/wing separation, but the gradient and position angle are robust.}
\label{tab:grads}
\begin{center}
\begin{tabular}{lcccc}
\hline
Southern boundary &  N &  $\overline{v}$ &  Gradient &  PA \\
 & & [\kms] & [\kms~deg$^{-1}$] & [deg] \\
\hline
$\delta = -77^{\circ}50^{\prime} + (4\alpha)^{\prime}$ & 1572 & 170.8 & $26.3{\pm}1.6$ & $126.2{\pm}3.9$ \\
$\delta = -77^{\circ}20^{\prime} + (4\alpha)^{\prime}$ & 1364 & 167.7 & $24.1{\pm}1.9$ & $122.5{\pm}5.3$ \\
$\delta = -78^{\circ}20^{\prime} + (4\alpha)^{\prime}$ & 1749 & 173.4 & $25.7{\pm}1.4$ & $129.1{\pm}3.3$ \\
$\delta = -79^{\circ}50^{\prime} + (6\alpha)^{\prime}$ & 1479 & 169.6 & $27.2{\pm}1.7$ & $125.7{\pm}4.0$ \\
\hline
\end{tabular}
\end{center}
\end{table*}

The amplitude of global velocity changes in our data, $\sim$50~\kms,
is comfortably larger than any plausible source of error we have
identified.  However, although the H$\;$\1 results reported by
\citet{ss97} and \citet{stan99} also show more positive velocities in the wing than the
bar (i.e., to the SE), the neutral-gas velocity field 
displays an overall NE/SW dominant gradient.  
The apparent discrepancy between the NW/SE gradient found here and the
SW/NE gradient in the neutral-gas map can be attributed in large part
to the region of largest H$\;$\1 velocities lying outside our survey
region, around 1h~20m, $-71^\circ 30\arcmin$ (Fig.~\ref{fig:stan}).
This region has a low density of stars (and gas), but purely
in terms of H$\;$\1\ velocity overwhelms the trend of increasing
velocity to the SE which dominates our data, and which is also present
in the H$\;$\1\ measures.

If we examine our results along the line of maximum gradient
identified by \citet[][their Fig.~3]{stan04} we reproduce their
results qualitatively, finding a gradient of 10~\kms~deg$^{-1}$ (along
a great circle at PA~=~60$^\circ$, through $\alpha = 0$h~45m, $\delta =
-73^\circ$; Fig.~\ref{fig:rotcurv}), though this still appears to be
somewhat smaller than the neutral-gas gradient.

\subsection{Induced gradients}
\label{sec:induced}

An extended source moving perpendicular to the line of sight at its
centre is not moving perpendicular to the line of sight elsewhere; the
line-of-sight projection of the transverse motion is therefore
non-zero other than at the centre.  Expressed in terms of observable
quantities, the resulting induced velocity gradient is
\[
8.273 \times 10^{-2}\,
\frac{\mu}{\mbox{mas yr}^{-1}}\,
\frac{D}{\mbox{kpc}}
\mbox{ km s}^{-1}\mbox{ deg}^{-1}
\]
for proper motion $\mu$ and distance $D$.

From observations with the \textit{Hubble Space Telescope (HST)},
\citet{kma06} determined a precise proper motion for the SMC.  Such is
the interest in their results that \citet{ppo08} have recently
reanalysed the {\it HST} data.  In Table~\ref{tab:pm} we list the
induced velocity gradients arising from these two measurements, and
also for the proper motion results of \citet{kb97} and \citet{mike99}.
In calculating the induced gradients we adopt $D = 60.6 \pm 1$~kpc
\citep{hhh05}.

The proper motions yield induced velocity gradients in the same
general direction as the gradient found in the 2dF results, but their
magnitudes are too small to explain our results.  Astrophysical and
induced gradients will add vectorially, so that the implied magnitude
of the residual astrophysical component of the early-type stars is
ca. 18~\kms.

\citet{hz06} found a velocity gradient for red giants of
8.3~\kms~deg$^{-1}$ -- about the same magnitude as the induced
gradient inferred here from the results of \citet{kma06}, but in an
almost orthogonal direction (PA~=~23$^\circ$; that is, roughly along the
major axis of the bar).  Independent data are required to establish if
the absence of induced gradients in red-giant velocities is a result
of errors in the measurements (velocities and/or proper motions) or is
astrophysical (implying rotation of the red-giant population about the
major axis, in the opposite sense to that found here, to compensate
geometrical effects).

\begin{table*}
\caption{Induced velocity gradients from published proper motion results.}
\label{tab:pm}
\begin{center}
\begin{tabular}{lcccc}
\hline
Source & $\mu_{\rm west}$ & $\mu_{\rm north}$ & $v_{\rm induced}$ & PA \\
& [mas yr$^{-1}$] & [mas yr$^{-1}$] & [$\mbox{ km s}^{-1}\mbox{deg}^{-1}$] & [deg]\\
\hline
\citet{kb97} & $-1.23 \pm 0.84$ & $-1.21 \pm 0.75$ & $8.7 \pm 5.6$ & $135 \pm 26$ \\
\citet{mike99} & $-0.92 \pm 0.20$ & $-0.69 \pm 0.20$ & $5.8 \pm 1.4$ & $127 \pm 10$ \\
\citet{kma06} & $-1.16 \pm 0.18$ & $-1.17 \pm 0.18 $ & $8.3 \pm 1.3$ & $135 \pm 6\phantom{1}$ \\
\citet{ppo08} & $-0.75 \pm 0.06$ & $-1.25 \pm 0.06$ & $7.3 \pm 0.4$ & $149 \pm 2\phantom{1}$ \\
\hline
\end{tabular}
\end{center}
\end{table*}

\begin{figure}
\begin{center}
\includegraphics[scale=0.6]{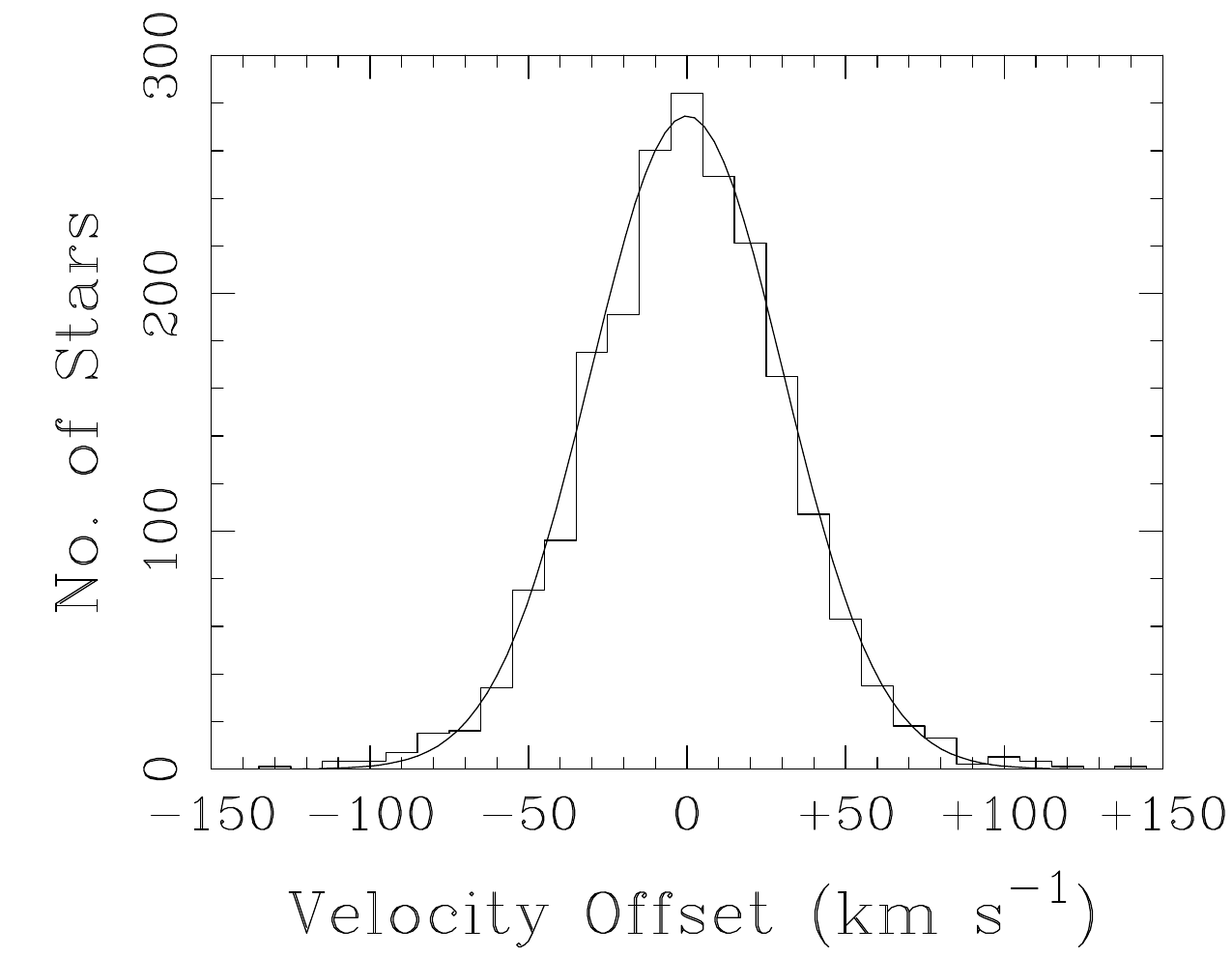}
\caption[]{Probability distribution function
of radial velocities after correction for systematic
gradients (cf.~Section~\ref{sec:vdisp2}).   A gaussian characterized
by $\sigma = 30.4$~\kms\ is shown for comparison.}
\label{fig:vrhist}
\end{center}
\end{figure}

\subsection{Velocity dispersions}
\label{sec:vdisp2}

If we take out the large-scale velocity trends in the data by
subtracting a simple quadratic fit from the results shown in
Fig.~\ref{fig:rotcurv} (left-hand panel), the weighted 1-$\sigma$
dispersion of the full set of velocity measurements is reduced from
33.6~\kms\/ to 30.4~\kms.  Weighted dispersions for `wing' and `bar'
stars are 28.6 and 33.3~\kms, respectively; subtracting a linear trend
from the bar sample reduces the latter figure to 30.7~\kms, which is
not significantly different from the wing value (as judged by an
$F$~test).

Adopting 30.4~\kms as the most representative figure for the observed
velocity dispersion, then, assuming that observational and
astrophysical contributions sum quadratically, the intrinsic
astrophysical velocity dispersion is ca. $\sim$28~\kms.  Although
Fig.~\ref{fig:rotcurv} shows that the total velocity range across the
observed area of the SMC is substantially larger than this velocity
dispersion, most probably that range reflects tidal effects, not
rotation.  Thus while our observational result contrasts in this
respect with \citeauthor{hz06}'s result for red giants, we concur that
the system is primarily supported by the velocity dispersion.

Interestingly, the velocity dispersion for the O-type spectra
(Table~\ref{tab:vels}) is found to be larger than that for the BA stars
(at greater than 99$\%$ confidence).  Weak nebular and/or wind
emission could contribute to this, but only one of the 83 O-type stars
with measured velocities is commented as having weak emission at
H$\gamma$ (cf. Table~A1 from Paper~I).  It is plausible that undetected
emission could be affecting the results, however Balmer emission is as
common in the B-type 2dF spectra as in the O-type stars (Table~7 from
Paper~I), so this does not suggest an explanation for the different
dispersions.  We also note that the great majority of the O-type stars
are either dwarfs or giants, in which wind effects in the higher-order
Balmer lines will be relatively minor.

We suggest that the larger dispersion likely originates from
undetected massive binaries and runaway stars.  The latter suggestion
relates to the interesting issue of the relative velocities of field
versus cluster stars \citep[aside from the separate issue that 4\% of
O-type stars appear to form in isolation,][]{dewit}.  Using the
catalogue of clusters and associations from \citet{bd00}, 18 of our
O-type stars with measured velocities are in the field \citep[as
defined using the 30~pc criterion from][]{pm95}.  For the 18 field
stars the (weighted) mean is $\overline{v} = 179.3 \pm 2.2$~\kms\/
($\sigma = 48.9$), as compared to $\overline{v} = 173.2 \pm
0.9$~\kms\/ ($\sigma = 39.3$) for the 65 others.  Unfortunately, with
such a limited sample, these results are not statistically
significant, but hint at a potentially interesting result, reinforcing
the case for a more optimised kinematic survey of the Clouds.

\section{Summary}

We have presented stellar radial velocities measured for 2045 stars
from our 2dF survey of the SMC.  We find a local velocity dispersion
that is approximately gaussian, with $\sigma \simeq 30$, similar to
recent results for red giants ($\sigma \simeq 27.5$~\kms;
\citealt{hz06}), but the global velocity range is significantly
larger, with the SMC wing receding some $\sim$20~\kms\ faster than the
bar.  We find a continuous distribution of velocities across the SMC,
but those in the wing appear to have a distinct distribution compared
to those in the bar (left-hand panel of Fig.~\ref{fig:rotcurv}),
presumably a consequence of the last close-interaction between the SMC
and LMC $\sim$0.2 Gyr ago \citep{mf80,yn03,ckg06}.  Within the bar we
find a mean gradient of $26.30 \pm 1.6$~\kms~deg$^{-1}$ at
PA~=~126$^\circ$ (cp. 8.3 ~\kms~deg$^{-1}$ at PA~$=$~23$^\circ$ for
the red giants), although, as the H$\;$\1 maps show, this simple
parameterization of the data almost certainly disguises a more complex
dynamical behaviour.

\section{Acknowledgements}
We thank Danny Lennon for the idle conversation on a Friday afternoon
at La Cuatro that led to us revisit the 2dF data.  We are also
grateful to Snezana Stanimirovi\'c for her kind permission to use
Fig.~1, and the referee for well-informed questions and
suggestions.  This paper is based on data obtained with the
Anglo-Australian Telescope.

\bibliography{evans_rv}

\bibliographystyle{mn2e}

\appendix
\section{Catalogue of stellar radial velocities}
\label{rvcat}

The full catalogue of radial velocities for the 2dF sample is available
through the on-line edition of Monthly Notices, and at
the Centre de Donn\'{e}es astronomiques de Strasbourg
(CDS)\footnote{\tt{ftp://cdsarc.u-strasbg.fr/pub/cats/J/MNRAS/XXX/YYY}}.

Table~\ref{catalogue} shows an extract from the catalogue, to
illustrate the format.  For each entry we give co\"ordinates, adopted
$B$ magnitudes (including their source) and spectral classifications,
all from Paper~I.  In the final three columns we give the mean and
standard error (Columns 7 \& 8, in \kms) of the measured velocities,
and the number of lines used (Column 9).

\begin{table*}
\begin{center}
\caption[] {An illustrative section of the radial-velocity catalogue.  Stellar radial velocities
($v_r$) and the standard error on the mean
($\sigma$/$\sqrt{n}$) are given in units of \kms; the final column reports the
number of Balmer lines that were measured for each star.  Sources of
adopted $B$ magnitudes are \citet{pm02}, OGLE \citep{u98},
\citet{z02}, and 2dF (i.e. values from the scanned Schmidt plates, see
Paper I), coded M, O, Z, and 2 respectively.}
\label{catalogue}
\begin{tabular}{ccccclccc}
\hline
2dFS\# & $\alpha$ & $\delta$ & $B$ & Source & Spectral type & $v_r$ & $\sigma/\sqrt{n}$ & $n$ \\
& (J2000) & (J2000) &&&& [\kms] & [\kms] & \\
\hline
 0600 & 00 46 23.13 & $-$72 50 17.4 &  16.36 &  O &   A3$\;$II      &  177 & 19 & 4 \\
 0606 & 00 46 33.47 & $-$73 39 25.8 &  13.34 &  M &   B8$\;$(Ib)    &  203 & 11 & 5 \\
 0609 & 00 46 38.10 & $-$73 55 13.7 &  14.31 &  Z &   B0.5$\;$(IV)  &  181 & 38 & 3 \\
 0620 & 00 46 50.88 & $-$73 55 21.5 &  14.57 &  Z &   B2$\;$(III)   &  210 & 14 & 5 \\
 0625 & 00 46 55.78 & $-$73 19 56.9 &  14.32 &  Z &   B9 (Ib)        & 130 & $\phantom{1}$7 & 5 \\
 0628 & 00 46 57.87 & $-$73 15 40.0 &  20.24 &  O &   B1-5 (V) & 151 & $\phantom{1}$8 & 4 \\
 0629 & 00 47 01.39 & $-$74 08 33.6 &  15.21 &  Z &   A2 II    & 186 & 14 & 4 \\
 0630 & 00 47 01.70 & $-$72 35 05.2 &  15.03 &  Z &   A0 (Ib)  & 180 & $\phantom{1}$4 & 5 \\
 0636 & 00 47 12.95 & $-$73 53 33.1 &  14.71 &  Z &   B2 (III) & 119 & $\phantom{1}$7 & 5 \\
 0637 & 00 47 14.46 & $-$73 25 39.6 &  15.40 &  Z &   B0-5 (III) & 145 & 20 & 4 \\
 0644 & 00 47 23.47 & $-$73 33 35.0 &  14.58 &  O &   B0-5 (II) & 165 & 26 & 3 \\
 0646 & 00 47 24.04 & $-$72 37 37.3 &  14.46 &  Z &   A5 II    & 155 & 13 & 4 \\
 0648 & 00 47 26.78 & $-$73 32 03.4 &  16.53 &  Z &   A3 II    & 199 & 16 & 3 \\
\hline
\end{tabular}                                                                         
\end{center}
\end{table*}

\end{document}